\documentclass[conference]{IEEEtran}
\usepackage{amssymb}
\usepackage{subfigure}
\usepackage[normalem]{ulem}
\usepackage[fleqn]{amsmath}
\usepackage[varg]{txfonts}
\usepackage{color}
\usepackage{multirow}
\usepackage{lscape}
\usepackage[ruled,linesnumbered]{algorithm2e}
\definecolor{pink}{rgb}{1,0.5,0.5}
\usepackage{amsmath}
\usepackage{amsfonts}
\usepackage{amssymb}
\usepackage{varwidth}
\usepackage{enumitem}
\usepackage{tikz}
\usetikzlibrary{calc}
\usetikzlibrary{positioning}
\usetikzlibrary{arrows}
\usetikzlibrary{decorations.pathreplacing}
\usetikzlibrary{fit}
\usetikzlibrary{matrix}
\tikzset{
    vertex/.style = {
        circle,
        fill            = black,
        outer sep = 2pt,
        inner sep = 1pt,
    },
    boxnode/.style = {align=center,draw}
}
\tikzstyle{arrow}=[draw, -latex,solid,line width=0.5pt]
\graphicspath{{./figures/}}
\makeatletter
\def\BState{\State\hskip-\ALG@thistlm}
\makeatother
\begin{document}
\title{Soft-Error and Hard-fault Tolerant Architecture and Routing Algorithm for Reliable 3D-NoC Systems}
\author{ \IEEEauthorblockN{Khanh N. Dang, Yuichi Okuyama, and Abderazek Ben Abdallah}
\IEEEauthorblockA{The University of Aizu \\
    Graduate School of Computer Science and Engineering \\
    Aizu-Wakamatsu 965-8580, Japan \\
    Email: \{d8162103,  okuyama, benab\}@u-aizu.ac.jp}
}
\maketitle

\begin{abstract}

Network-on-Chip (NoC) paradigm has been proposed as an auspicious solution to handle the strict communication requirements between the increasingly large number of cores on a single multi and many-core chips. However, NoC systems are exposed to a variety of manufacturing, design and energetic particles factors making them vulnerable to permanent (hard) faults and transient (soft) errors. 
In this paper, we present a comprehensive soft error and hard fault tolerant 3D-NoC architecture, named 3D-Hard-Fault-Soft-Error-Tolerant-OASIS-NoC (3D-FETO). With the aid of adaptive algorithms, 3D-FETO is capable of detecting and recovering from soft errors occurring in the routing pipeline stages and is leveraging on reconfigurable components to handle permanent faults occurrence in links, input buffers, and crossbar. In-depth evaluation results show that the  3D-FETO system is able to work around different kinds of hard faults and soft errors while ensuring graceful performance degradation, minimizing the additional hardware complexity and remaining power-efficient. 

\end{abstract}
\begin{IEEEkeywords}
    Fault Tolerance, Routing Algorithm, 3D Network-on-Chip, Architecture
\end{IEEEkeywords}
\IEEEpeerreviewmaketitle

\section{Introduction}

In the past few years, the benefits of 3D Integrated Circuits (3D-ICs) and the regularity of mesh-based Network-on-Chips (NoCs) have been fused into a promising architecture, called 3D-Network-on-Chip (3D-NoC)~\cite{BenAhmed2013Architecture}, opening a new horizon for IC design. In fact, the parallelism of Network-on-Chip can be enhanced in the third dimension thanks to the short wire length and low power interconnects of 3D-ICs. As a result, the 3D-NoC paradigm is considered as one of the most advanced and auspicious architectures for the future of IC design, as it is capable of providing extremely high bandwidth and low power interconnects.

While the NoC paradigm has been increasing in popularity with several commercial chips, 
it is threatened by the decreasing reliability of aggressively scaled transistors. Transistors are approaching the fundamental limits of scaling, with gate widths nearing the molecular scale, resulting in break down and wear out in end products.  
Therefore, future complex 3D-NoCs systems will require significant tolerance to many simultaneous soft errors and hard faults.

Hard faults, including both permanent faults and intermittent faults, can occur during the manufacturing stage or under specific operation circumstances.
For both permanent and intermittent faults, the most natural solution is using redundant components.

Soft errors arise from energetic particles such as alpha particles and neutrons from cosmic rays generating electron-hole pairs as they pass through a device. 
Soft errors do not permanently defect the gate and only occur in a short period of time. Because of their special characteristics, they are unpredictable and unavoidable. Unlike permanent and intermittent faults, transient faults cannot be fixed by just replacing the affected component. Instead, they can be recovered from by repeating the erroneous operation or information redundancy (e.g., Error Correction Code (ECC)). 
Therefore, without efficient protection mechanism, these errors can compromise the system's functionality and reliability.

Most of the conducted works handle the hard faults and soft errors separately. Hard faults handling schemes are mainly based on two main approaches: (a) fault-tolerant routing algorithms which enable packets to avoid faulty nodes in the network~\cite{BenAhmed2013Architecture,DeOrio2012reliable} and (b) architecture-based methods which use hardware (components) redundancy or/and reconfiguration to recover from faults~\cite{DeOrio2012reliable,Constantinides2006Bulletproof,BenAhmed2016Adaptive}. 
Soft errors recovery is  also solved by two main schemes: (a) Error Correction Code (ECC) based methods for data corruption~\cite{lin1984arr,Bertozzi2005Error,yu2010transient} and (b) Control logic (temporal redundancy) based methods ~\cite{Ernst2003Razor,	 Yu2013Addressing,Dang2015Softa}. 
\\
Although theses works provide solutions to separately handle hard faults and soft errors in the router, no comprehensive solutions were proposed to simultaneously handle both hard faults and soft errors in a 3D-NoC system. In addition, the error  detection  and  diagnosis in a NoC  architectures  have  been studied thoroughly in the scope of offline testing. However, with soft errors and intermittent faults becoming a dominant failure mode in modern NoC and general VLSI systems, a widespread deployment of online test  approaches has become crucial. 
 
In this paper, we present a comprehensive soft error and hard fault tolerant 3D-NoC architecture, named 
3D-Hard-Fault-Soft-Error-Tolerant-OASIS-NoC (3D-FETO). 
The main contributions of this work are summarized as follows: 
\begin{figure*}[bhtp]
    \centering
    \includegraphics[width=0.84\linewidth]{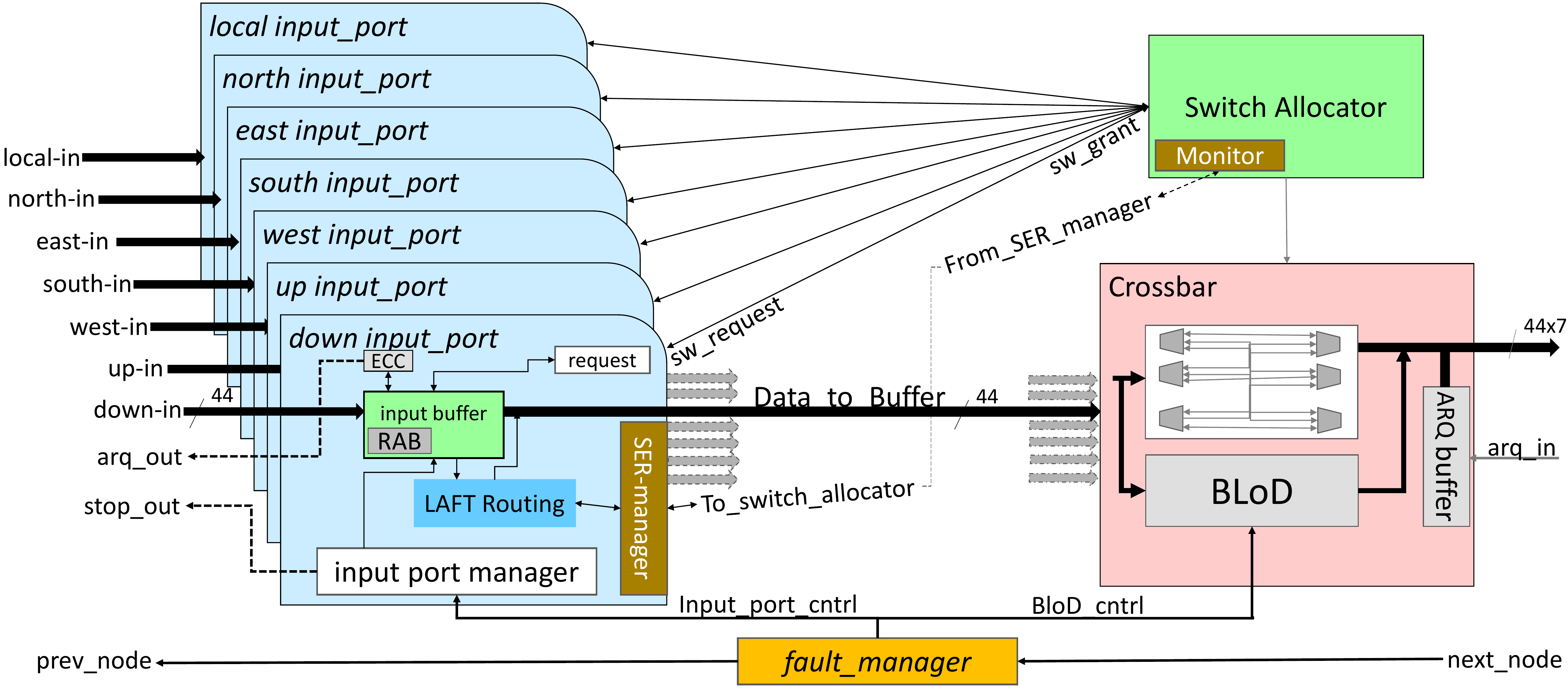}
    \caption{Adaptive 3D router (SHER-3DR) architecture.}
    \label{fig:R_A_2}
\end{figure*}  
\begin{itemize}
    \item New fault-tolerant routing algorithm and architecture which allows the system to handle soft errors and hard faults at the same time. 
    \item An  efficient  scheme  for  online  control  fault  detection  and  diagnosis  in  3D-NoC  systems. 
    \item Quantitative evaluation and analysis of the proposed system with different synthetic and realistic benchmarks. 
\end{itemize}

\section{Adaptive 3D Router Architecture (SHER-3DR)} \label{sec:arch}

Figure~\ref{fig:R_A_2} shows the block diagram of the proposed adaptive 3D router architecture (SHER-3DR). The router relies on simple recovery techniques based on system reconfiguration with redundant structural resources to contain hard faults in input-buffers, crossbar, and links, in addition to soft errors in the routing pipeline stages. 

The SHER-3DR router is the backbone component of the 3D-FETO system. Each router has a maximum number of 7-input and 7-output ports, where 6 input/output ports are dedicated to the connection to the neighboring routers and one input/output port is used to connect the switch to the local computation tile. As shown in Fig.~\ref{fig:R_A_2}, the SHER-3DR contains seven \textit{Input-port} modules for each direction in addition to the \textit{Switch-Allocator}, and the \textit{Crossbar} module which handles the transfer of flits to the next neighboring node.  
There are three pipeline stages on the router: Buffer-Writing (storing the incoming flits), Next-Port-Computing/Switch-Allocator (Routing and Arbitrating) and Crossbar-Traversal (Crossbar).

In this section, we first review the hard-fault handling mechanism  \cite{BenAhmed2016Adaptive}, including the Random-Access-Buffer (RAB) for deadlock-recovery fault-tolerance, and the Bypass-Link-on-Demand (BLoD) approach to handle multiple faulty channels in the crossbar. Secondly, we present a soft error recovery mechanism. In order to support detection and recovery, a light-weight detection, diagnosis and recovery is depicted in the last subsection.
\subsection{Hard Fault Recovery Mechanism Overview}  \label{ssec:HER}

The hard fault recovery mechanisms~\cite{BenAhmed2016Adaptive} consist of Random Access Buffer and Bypass-Link-on-Demand. The Random Access Buffer mechanism (RAB)~\cite{BenAhmed2016Adaptive} solves the deadlock problem that can occur with the look-ahead fault-tolerant routing algorithm (LAFT), and is able to recover from intermittent and permanent faults in the input-buffer. When a fault is detected in one of the slots, the main controller 
takes into consideration the flagged slots when assigning the write and read addresses. When a slot is marked as faulty, it will be avoided in reading and writing processes. By using this mechanism, RAB can handle the presence of the faults in input-buffer.

The Bypass Link on Demand mechanism (BLoD) \cite{BenAhmed2016Adaptive} provides additional escape channels whenever the number of faults in the baseline 7x7 crossbar increases. In the case where a fault is detected in one or several crossbar links, the BLoD's controller 
disables the faulty crossbar links and enables the appropriate number of bypass channels.
The number of Bypass-links is very important and it should be minimized as much as possible to reduce the area and power overhead. In the case where the number of faulty links are larger than the number of backup links, the system needs to mark the router-to-router connection as faulty and use the LAFT algorithm to avoid it~\cite{BenAhmed2016Adaptive}.

\subsection{Soft Error Recovery Mechanism}

The principle of soft-error handling method in 3D-FTO relies on duplicating the pipeline stages computation (software redundancy) in one more clock cycle, and detection is made based on the difference between the computed results. If two clock cycles have similar results, there is no soft-error.  If two consecutive results are different, there is a soft-error. In this case, the system requires a third clock-cycle in order to correct the failure. The failure is corrected by majority voting of three results. To recover from soft errors in the data, the conventional ECC (Error Correction Code)~\cite{Bertozzi2005Error, Yu2012TPE} is adopted.

\begin{figure}[htbp]
    \centering
    \includegraphics[width=1\linewidth]{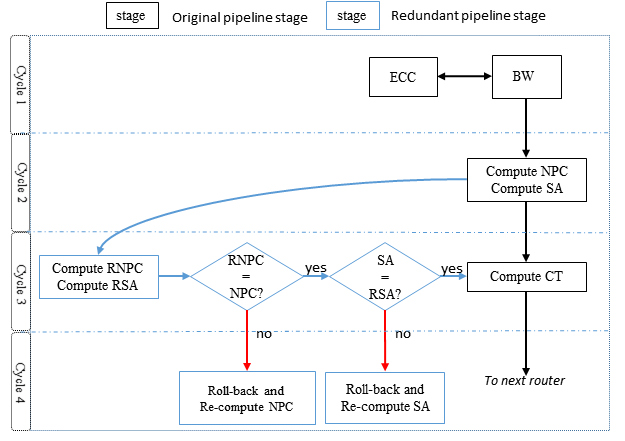}
    \caption{SHER-3DR Pipeline Stages Time-chart.}
    \label{fig:R_PP}
\end{figure}

For ease of understanding, we provide the router pipeline stages time-chart in Fig.~\ref{fig:R_PP}. The Next Port Computation (NPC) and Switch Allocation (SA) run in parallel as shown in \textit{Cycle 2} of Fig.~\ref{fig:R_PP}. This is achieved by the LAFT routing algorithm (described later), where the dependency between the two stages is eliminated. The duplication is made for NPC, and SA stages. After the first computation (in \textit{Cycle 2}), all the these stages have an additional computation clock. If a soft error is detected, the whole pipeline is halted for correction. 

In \textit{Cycle 1}, flits are stored in the input buffer at the Buffer Writing (BW) stage, and the ECC is used to check and correct the input data in the ECC module. In \textit{Cycle 2}, the NPC and the SA are executed in parallel in the LAFT routing unit. In \textit{Cycle 3}, the Redundant NPC (RNPC) and the Redundant SA (RSA) are computed in parallel. Then, if the output of RNPC is equal to that of NPC, and SA is equal to RSA, the Crossbar Traversal (CT) stage is performed in \textit{Cycle 3}, and the flit goes to the next router via the output channel in \textit{Cycle 4}. If the RNPC is not equal to the NPC, the system rolls-back and recomputes the NPC. Moreover, if SA is not equal to RSA, the system also rolls-back and re-computes the SA stage in \textit{Cycle 4}. The third results are attained from re-executing of the failed stages. The router can determine the correct result by using majority voting of three results.

\subsection{Light-weight Detection, Diagnosis and Recovery Mechanism (DDRM)} \label{ssec:DDRM}


Algorithm~\ref{alg:DRA} shows the the proposed \textit{Detection, Diagnosis and Recovery Mechanism} (DDRM) approach. It uses the feedback from ECC and the Automatic Retransmission Request (ARQ) protocol to monitor the errors. This mechanism is operated by the \textit{Fault-manager} module in Fig.~\ref{fig:R_A_2}.
As shown in Fig.~\ref{fig:R_A_2}, the input data is first verified by an ECC decoder.
If the value is correct or the ECC decoder can handle the correction, the flit is  written into the input buffer. Otherwise, a retransmission is requested. 
Since the transient fault only occurs in a short period of time, assumed to be a single clock cycle, it does not occur in two consecutive cycles. Therefore, ARQ can recover this kind of faults. However, if a permanent fault occurs, ARQ is unable to correct it and the faulty connection will keep the retransmission request infinitely. Therefore, if the ARQ cannot correct the fault, the system considers it as a permanent fault(line 1-10 in Algorithm~\ref{alg:DRA}).

Since the flit's correctness is verified by the ECC module before being written into the buffer, a permanent fault can only occur in the path between the input-buffer in the upstream node and the one in the downstream node. This path includes input buffers, crossbar and router-to-router channel.

For the diagnosis and recovery phase, the router's \textit{Fault-manager} module initiates the diagnosis with input buffer checking. In this step, the error status of the following flits of the monitored input buffer is checked. If errors are repeated at the same buffer position, \textit{Fault-manager} concludes this position is failed and sends information to \textit{Random Access Buffer} (RAB).  If errors are detected at another buffer position, the fault belong to crossbar or router-to-router link. In this case, the algorithm changes to check crossbar.

In the crossbar checking step, an alternative link by \textit{Bypass-Link-on-Demand} (BLoD) is selected for another flit. If this flit is healthy after its transmission, the bypass-link fixed the fault. The configuration of BLoD is keep as recovery. If this flit is still failed, the bypass-link was unable to fix it. The fault belong to the router-to-router channel. The fault information is sent to \textit{Look-Ahead Fault-Tolerant} routing modules to avoid the connection as recovery.

\section{Look-Ahead-Fault-Tolerant Routing Algorithm}\label{sec:LAFT}

To keep the benefits of look-ahead routing,  
Look-Ahead-Fault-Tolerant routing algorithm (LAFT) \cite{BenAhmed2013Architecture} should be able to perform the routing decision for the next node taking into consideration its link status and select the best minimal path. The fault link information received from the \textit{Fault\_manager} (which handles the {\it Detection Diagnosis Recovery Mechanism}) are read by each input-port where LAFT is executed. Algorithm~\ref{alg:LAFT} illustrates the LAFT algorithm. The first phase of this algorithm calculates the next node address depending on the \textit{Next-port} identifier read from the flit. For a given node wishing to send a flit to a given destination, there exist at most three possible directions through X, Y, and Z dimensions, respectively. In the second phase, LAFT performs the calculation of these three directions. 
By the end of this second phase, LAFT has information about the next node's fault status and also the three possible directions for a minimal routing. In the next phase, the routing selection is performed. For this decision, we adopted a set of prioritized conditions to ensure fault-tolerance and high performance either in the presence or absence of faults:
\begin{enumerate}
    \item The selected direction should ensure a minimal path and it is given the highest priority in the routing selection.
    \item We should select the direction with the largest next-hope path diversity.
    \item The congestion status is given the lowest priority. 
\end{enumerate}
By the end of the selection, a routing path with largest diversity and lowest congestion is selected. If there is no minimal routing path due to the presence of faults, another selection is applied for other non-minimal routing directions.

\begin{algorithm}[hbpt]
    \scriptsize
    \caption{Fault Detection, Diagnosis and Recovery.}
    \label{alg:DRA}
    \tcp{Automatic Retransmission Request}
    \KwIn{$transmitting\_flit$}
    \tcp{Transmitted Buffer Position}
    \KwIn{$buffer\_position$}
    
    \tcp{Control signal to all Fault-Tolerance modules}
    \KwOut{$RAB\_control$, $BLoD\_control$, $LAFT\_control$}
    
    
    \tcp{Transmit the flit, get the ECC's feedback}
    \textbf{Transmit}($transmitting\_flit$);
    
    $ECC\_result$ = \textbf{ECC-Decoder}($transmitting\_flit$);
    
    \tcp{DETECTION PHASE:}
    
    \uIf{$ECC\_result == ARQ$}{
        \tcp{Automatic Retransmission Request}
        \textbf{increase}($ARQ\_counter$);
        
        \textbf{ARQ}($transmitting\_flit$);
    } \Else {
    \tcp{The transmitted flit is non faulty}
    \textbf{Finish};
}
\tcp{Check the number of consecutive ARQs}
\If{($ARQ\_counter == 2$)}{
    \tcp{There is a permanent fault}
    \tcp{Jump to DIAGNOSIS-RECOVERY PHASE}
}

\tcp{DIAGNOSIS-RECOVERY PHASE:}
\tcp{Start with Input Buffer Checking}
$Buffer\_Failure \leftarrow Buffer\_Checking(buffer\_position)$;

\uIf{($Buffer\_Failure == Yes$)}{
    \tcp{Random Access Buffer is received the position to handle.}
    
    $RAB\_Control = buffer\_position$;
    
    \textbf{Finish};
}\Else{ 
\tcp{The buffer slot is non faulty.}
\tcp{Move to Crossbar Checking: using a Bypass-Link.}
$BLoD\_control$ = enable;

\tcp{Get the ECC's feedback and detect with ARQ counter.}

\uIf{($ARQ\_counter == 2$)}{
    \tcp{BLoD cannot fix the fault, the link is failed.}
    
    $BLoD\_control$ = release;
    
    \tcp{The LAFT routing algorithm handles the faulty link.}
    $LAFT\_control$ = faulty;
    
    \textbf{Finish};
}\Else{
\tcp{BLoD already fixed the failure, the recovery step is finished.}
\textbf{Finish};
}

}
\end{algorithm}
\begin{algorithm}[hbtp]
    \scriptsize
    \caption{Look-Ahead-Fault-Tolerant Routing.} \label{alg:LAFT}
    \tcp{Destination address}
    \KwIn{$X_{dest}$, $Y_{dest}$, $Z_{dest}$}
    \tcp{Current node address}
    \KwIn{$X_{cur}$, $Y_{cur}$, $Z_{cur}$}
    \tcp{Next-port identifier}
    \KwIn{{\it Next-port}}
    \tcp{Link status information}
    \KwIn{Fault-in}
    \tcp{New-next-port for next node}
    \KwOut{{\it New-next-port}}
    
    \tcp{Calculate the next-node address}
    {\it Next}$\leftarrow$ {Next-node} ({\it $X_{cur}$, $Y_{cur}$, $Z_{cur}$, Next-port});
    
    \tcp{Read fault information for the next-node}
    {\it Next-fault}$\leftarrow$ {Next-status} ({\it Fault-in, Next-port});
    
    \tcp{Calculate the three possible directions for the next-node}
    {\it Next-dir}$\leftarrow$ {poss-dir} ({\it $X_{dest}$, $Y_{dest}$, $Z_{dest}$, $Next_x$, $Next_y$, $Next_z$});
    
    \tcp{Evaluate the diversity number of three minimal paths}
    
    $Div_{1}$ $\leftarrow$ {path-div} ({\it $X_{dest}$, $Y_{dest}$, $Z_{dest}$}, $poss-dir_1$);
    
    $Div_{2}$ $\leftarrow$ {path-div} ({\it $X_{dest}$, $Y_{dest}$, $Z_{dest}$}, $poss-dir_2$);
    
    $Div_{3}$ $\leftarrow$ {path-div} ({\it $X_{dest}$, $Y_{dest}$, $Z_{dest}$}, $poss-dir_3$);
    
    \tcp{Evaluate the New-next-port direction}
    \uIf{($|$Next-dir$|$ $>$ 1)}
    {\uIf {($Div_{1}$==$Div_{2}$==$Div_{3}$)}
        {
            {\it New-next-port} $\leftarrow$ {min-congestion ($poss-dir_{1}$, $poss-dir_{2}$, $poss-dir_{3}$)};
        }
        
        \Else {
            {\it New-next-port} $\leftarrow$ {max-diversity  ($poss-dir_{1}$, $poss-dir_{2}$, $poss-dir_{3}$)};
        }
    }
    \Else { \uIf{(Next-dir == 1)}
        {{\it New-next-port}$\leftarrow$ $Next-dir_1$;}
        \lElse{{\it New-next-port}$\leftarrow$ {nonminimal} ({\it $X_{dest}$, $Y_{dest}$, $Z_{dest}$, $X_{cur}$, $Y_{cur}$, $Z_{cur}$, Fault-in})}
    }   
\end{algorithm}                             

\section{Evaluation Results}\label{sec:eva}

%
The proposed 3D-FETO system was designed in Verilog-HDL, synthesized and prototyped with commercial CAD tools and VLSI technology, respectively \cite{NCSUEDA2015FreePDK3D45,nangate2014nangate}. We evaluate the hardware complexity of 3D-FETO router in terms of area utilization, power consumption (static and dynamic) and speed. To evaluate the performance of the proposed system, we select both synthetic and realistic traffic patterns as benchmarks. For synthetic benchmarks, we select Transpose , Uniform, Matrix-multiplication, and Hotspot 10\%. For realistic benchmarks, we choose traffic patterns of H.264 video encoding system, Video Object Plane Decoder (VOPD), Picture In Picture (PIP) and Multiple Window Display (MWD)~\cite{Bertozzi2005NoC}. The simulation configurations are depicted in Table~\ref{tab:sim-conf}.

We evaluate the performance of our fault-tolerant model which includes hard fault tolerance from 3D-FTO~\cite{BenAhmed2016Adaptive}, soft-error resilience (SER-OASIS), and the proposed system (3D-FETO). We measure the network transmission time, or end-to-end latency, with the selected synthetic and realistic benchmarks. To understand the impact of fault-tolerant techniques on performance, we compare the obtained results with the baseline 3D-NoC system presented in~\cite{BenAhmed2013Architecture}. We randomly inject faults with three fault-rates: 10\%, 20\% and 33\%.

\subsection{Latency Evaluation}

In the first experiment, we evaluate the performance of the proposed architecture in terms of latency over various benchmark programs and error injection rates for three system configurations: (1) hard-fault tolerant system (3D-FTO), (2) soft-error resilience system (Soft Error Resilience OASIS), and (3) Hard-fault and Soft-error tolerant system (3D-FETO). The simulation results are shown in Fig.~\ref{fig:all-results}.(a-h). From these graphs, we notice that with 0\% hard faults (in input buffer and crossbar only), 3D-FTO has almost similar performance as the baseline system (LAFT-OASIS). In addition, we found that even at 33\% fault-rate, 3D-FTO increases the latency by only 1.71\%, 11.38\%, 8.79\% and 13.73\% for Transpose, Uniform, $6\times 6$ Matrix, and Hotspot-10\%, respectively. 
With realistic benchmarks, the performance of 3D-FTO slightly degrades at low error-rates; but, it suffers from more impact with high error-rates (20\% and 33\%) since the important connectivity encounters bottlenecks due to errors inside the input buffers. However, the proposed 3D-FETO model is still working even at high fault-rates while the baseline model collapses even at 5\% error-rate. \\
We used the same benchmark programs to evaluate the soft error resilience (SER) model. Since both of proposed \textit{Soft Error Resilience} and ECC require additional clock cycles, we can observe the significant effect on network transmission time. With 0\%, 10\%, 20\% and 33\% of fault-rates, the SER model increases the average delay in Transpose benchmarks by 18.57\%, 28.74\%, 34.54\% and 49.62\%, respectively. 

Finally, we evaluate the proposed 3D-FETO system with both soft error and hard faults handling schemes. As shown in Fig.~\ref{fig:all-results}(a-h), 3D-FETO has demonstrated a significant impact on the average latency which is mostly doubled in both realistic and synthetic benchmarks. At 33\% of fault-rates in Matrix, Uniform, Transpose benchmarks, 3D-FETO's average latency increases by 78.44\%, 50.73\% and 67.18\% in terms of average packet latency. However, it still maintains the ability of working under an extremely high fault-rate (33\% of hard faults and 33\% of soft errors).

\subsection{Throughput Evaluation}

\begin{figure*}[bht]
    \centering
    \subfigure[Transpose's Average Packet Latency]{\includegraphics[width=0.3\linewidth]{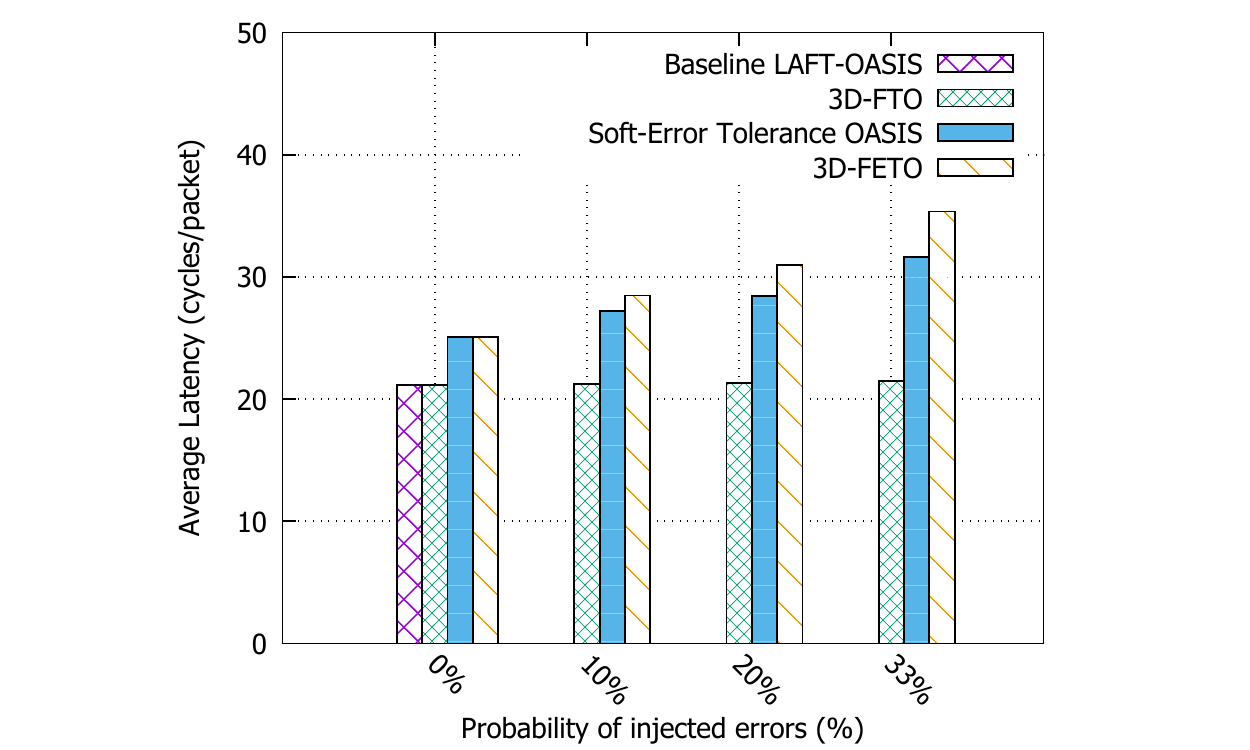}}
    \subfigure[Uniform's Average Packet Latency]{\includegraphics[width=0.3\linewidth]{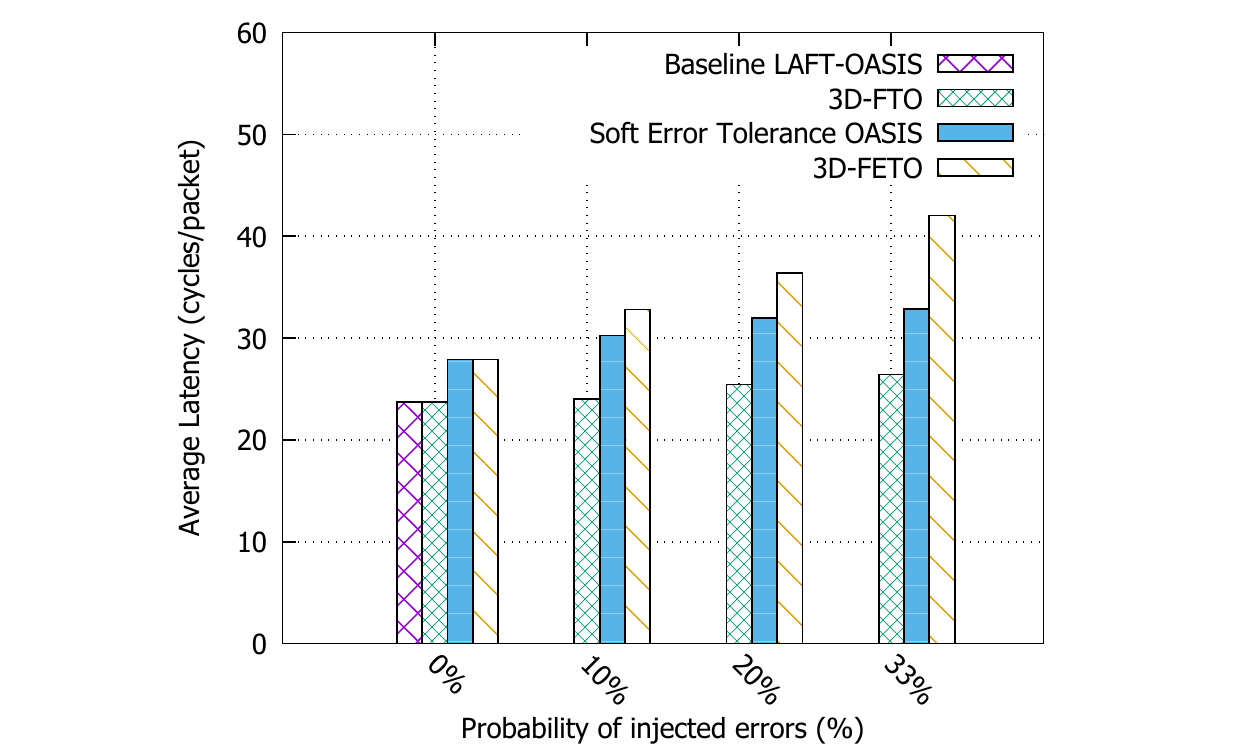}}   
    \subfigure[6 $\times$ 6 Matrix's Average Packet Latency]{\includegraphics[width=0.3\linewidth]{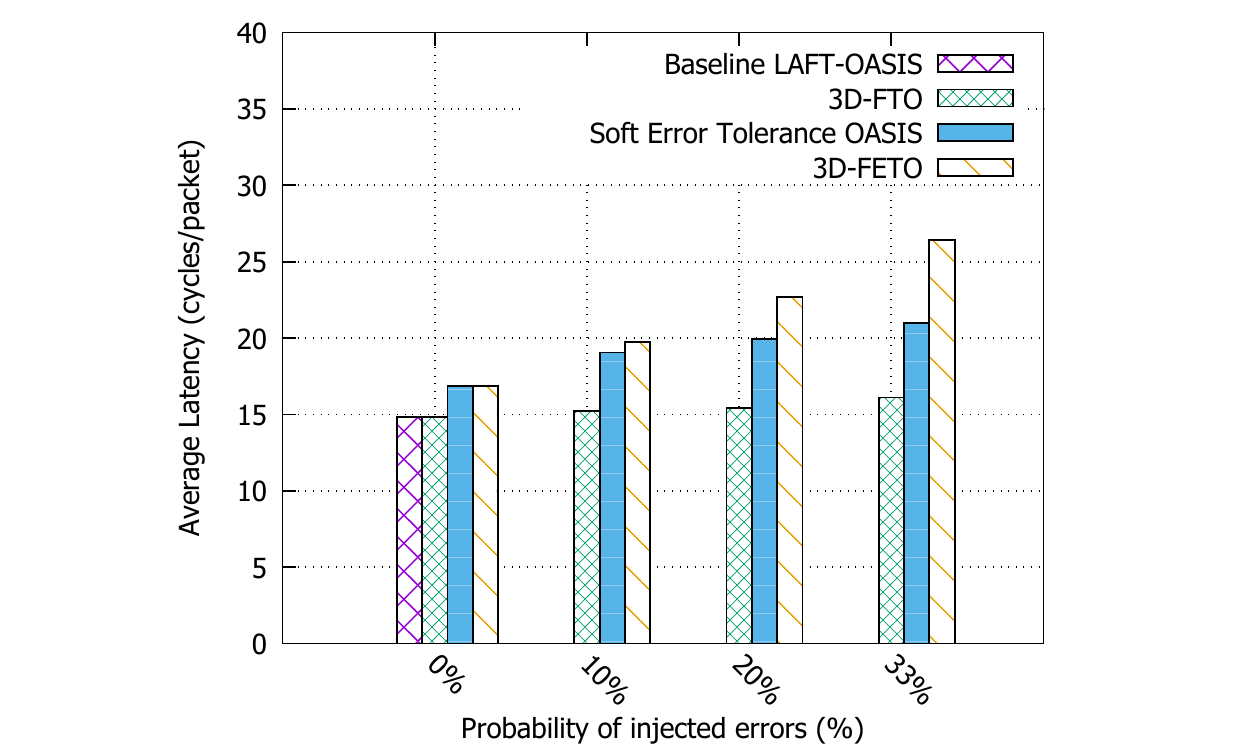}}\\
    \subfigure[Hotspot 10\%'s Average Packet Latency]{\includegraphics[width=0.3\linewidth]{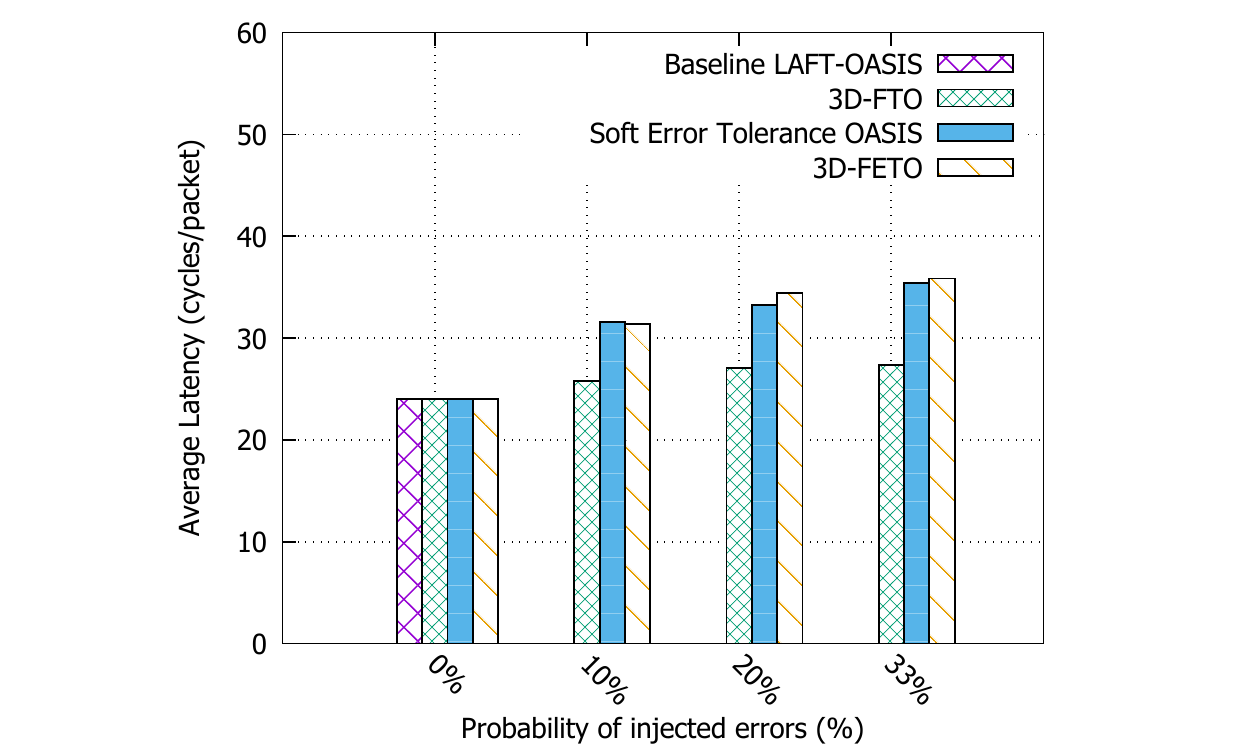}} 
    \subfigure[H.264 Encoder's Average Packet Latency]{\includegraphics[width=0.3\linewidth]{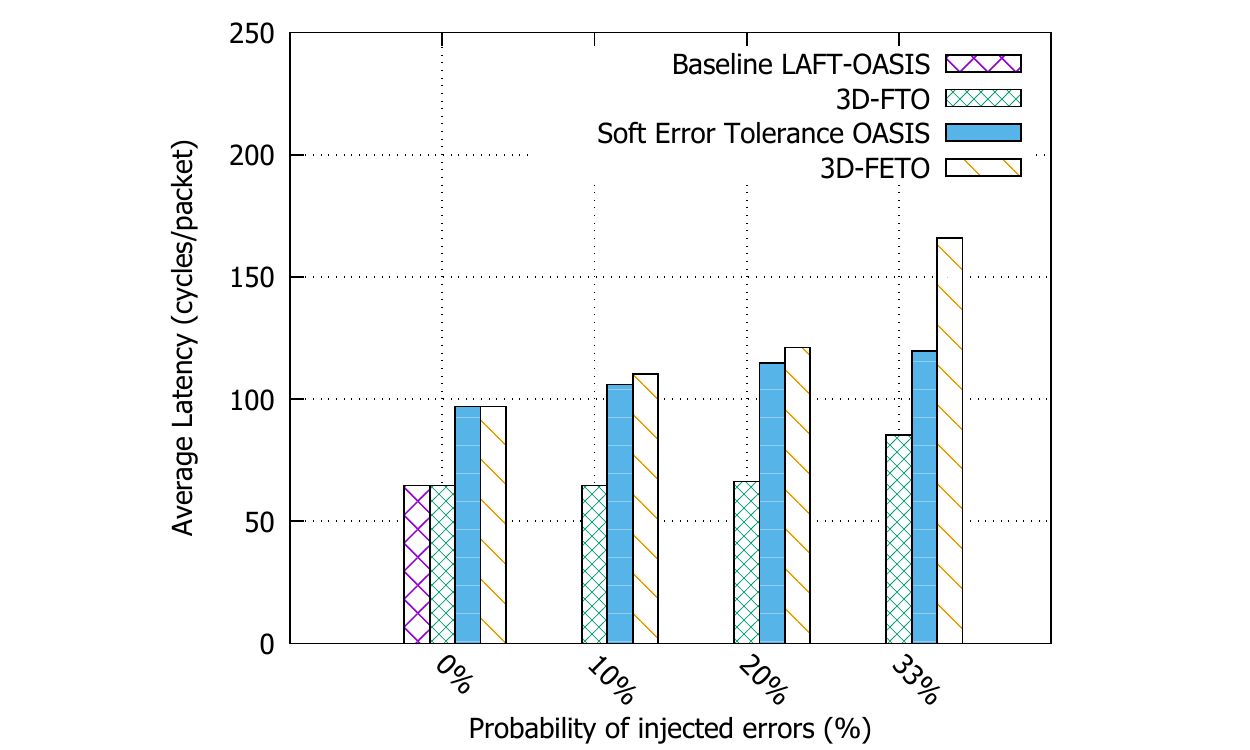}}   
    \subfigure[VOPD's Average Packet Latency]{\includegraphics[width=0.3\linewidth]{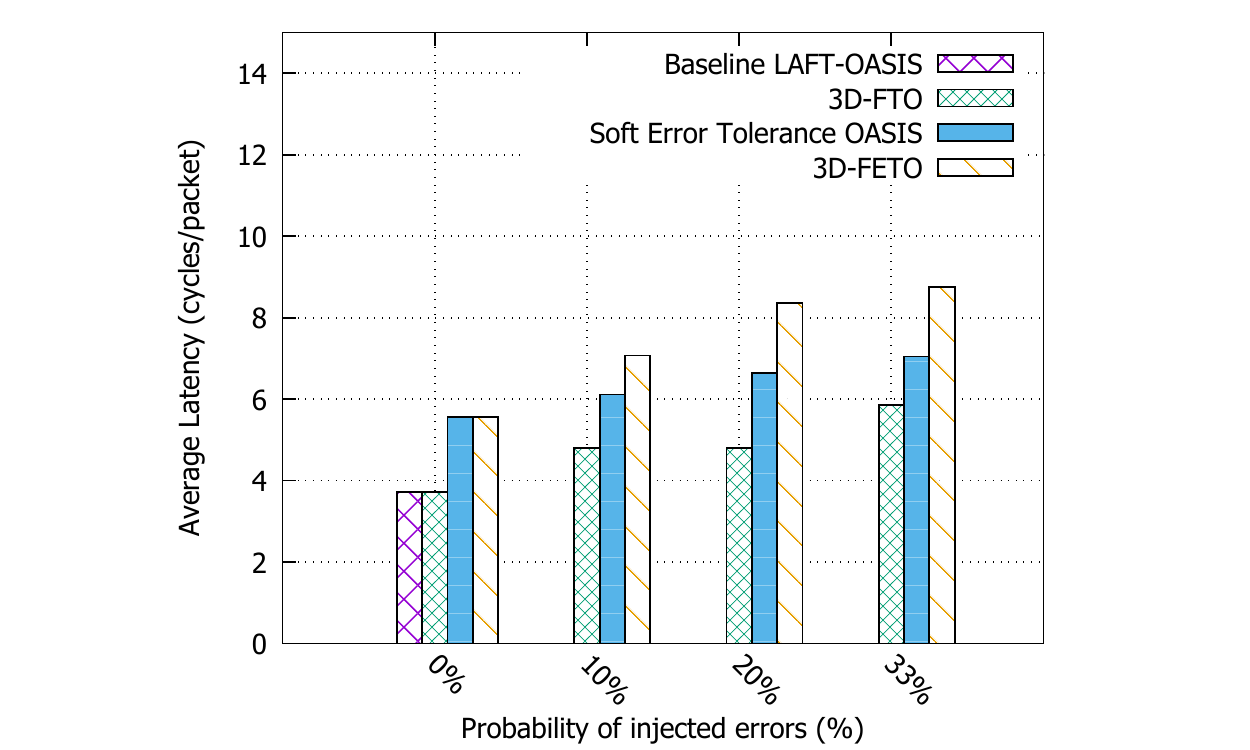}}\\  
    \subfigure[MWD's Average Packet Latency]{\includegraphics[width=0.3\linewidth]{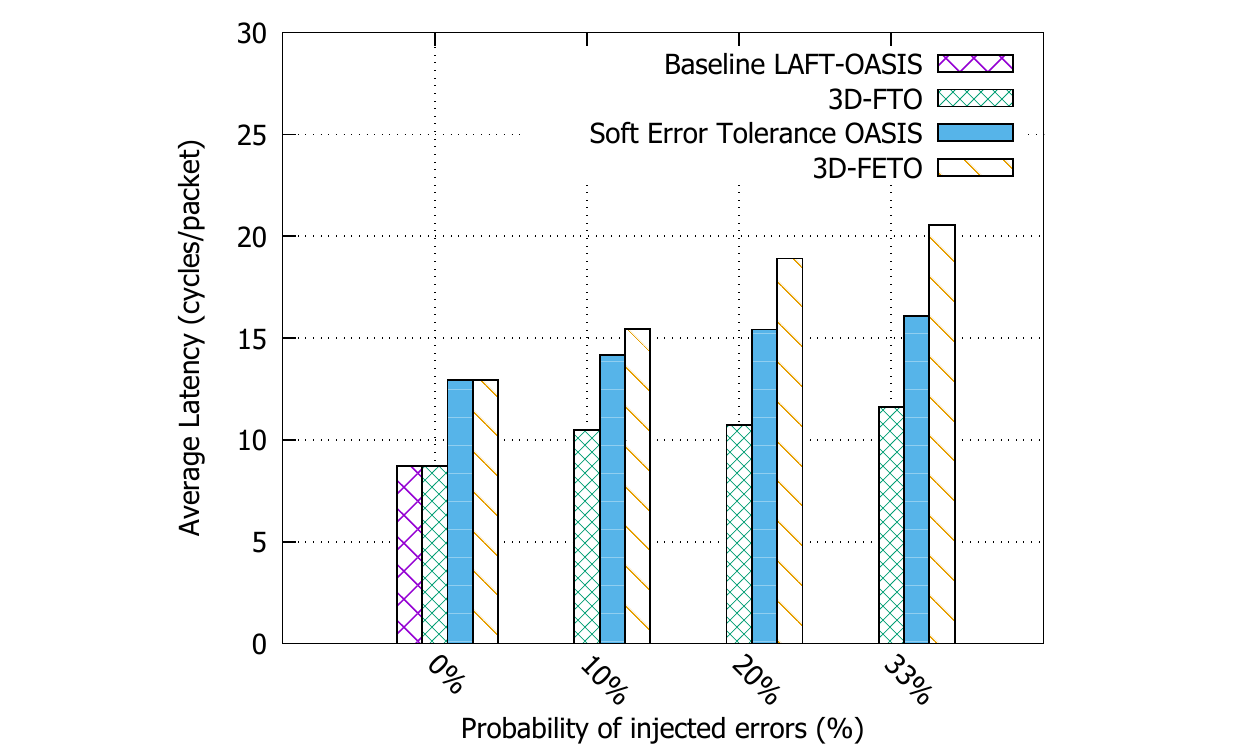}}   
    \subfigure[PIP's Average Packet Latency]{\includegraphics[width=0.3\linewidth]{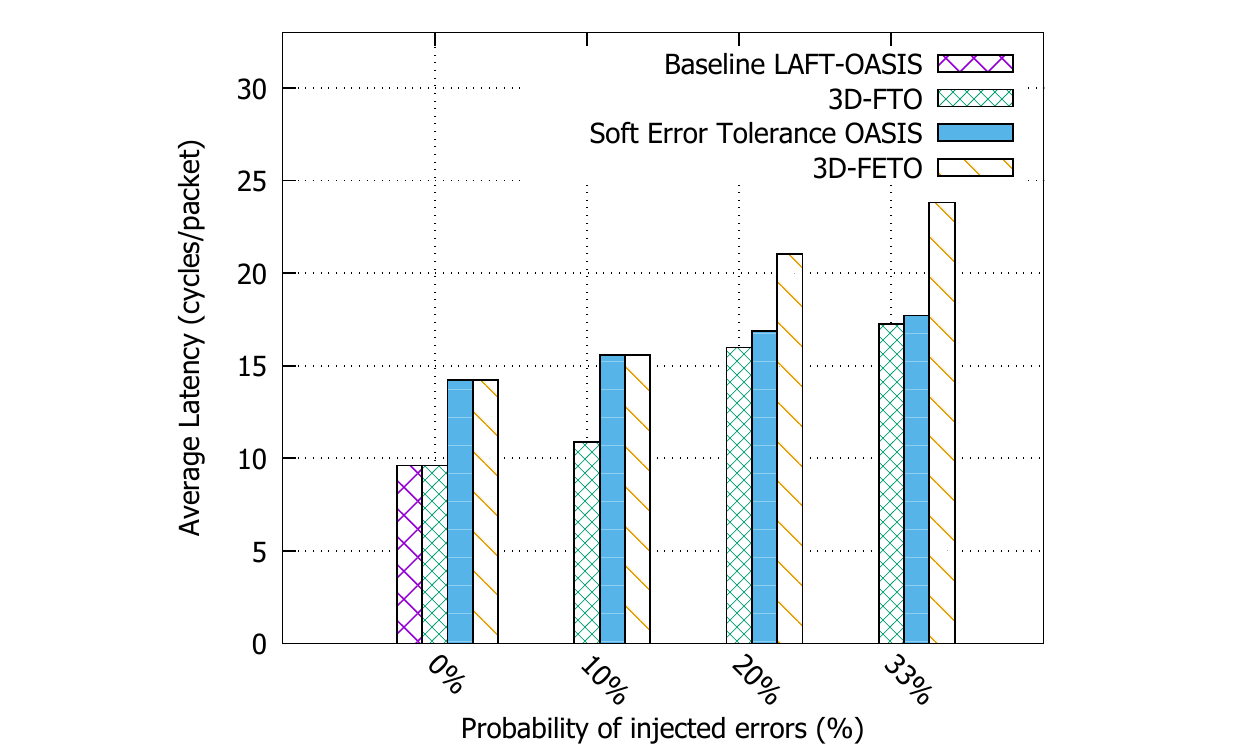}}
    \subfigure[Transpose's Throughput]{\includegraphics[width=0.3\linewidth]{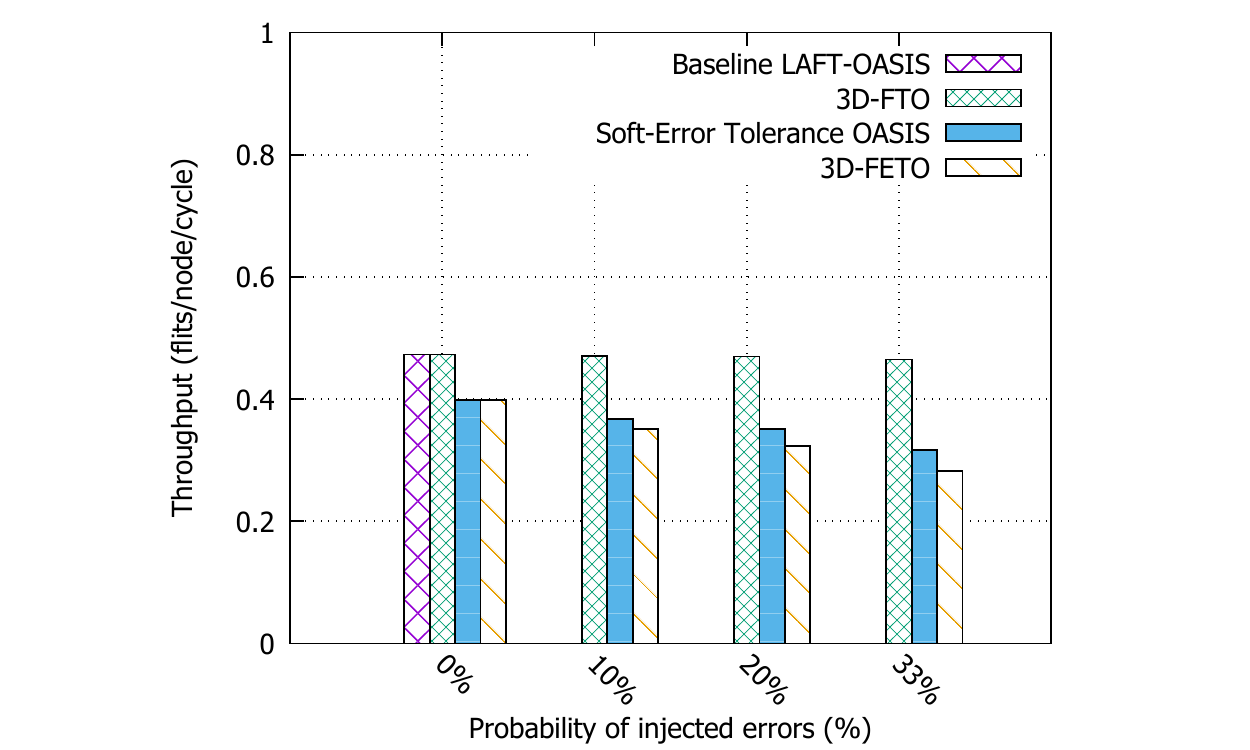}}\\
    \subfigure[Uniform's Throughput]{\includegraphics[width=0.3\linewidth]{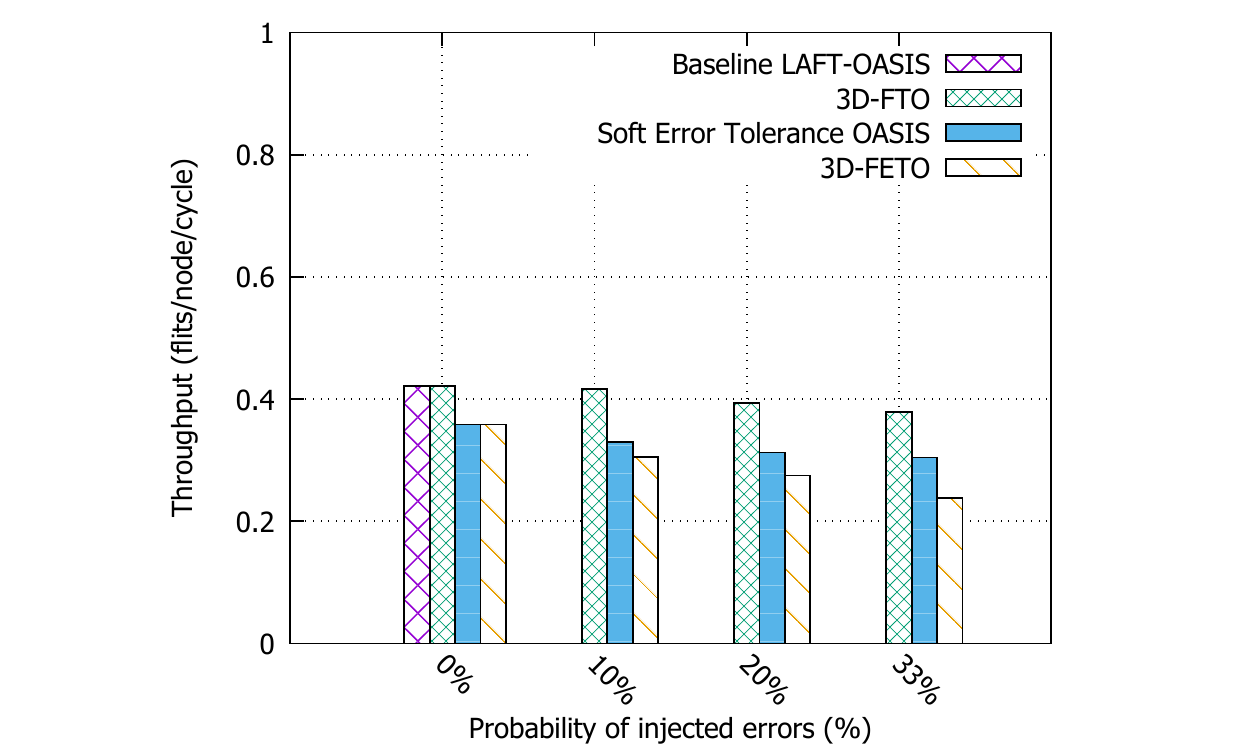}}   
    \subfigure[6 $\times$ 6 Matrix's Throughput]{\includegraphics[width=0.3\linewidth]{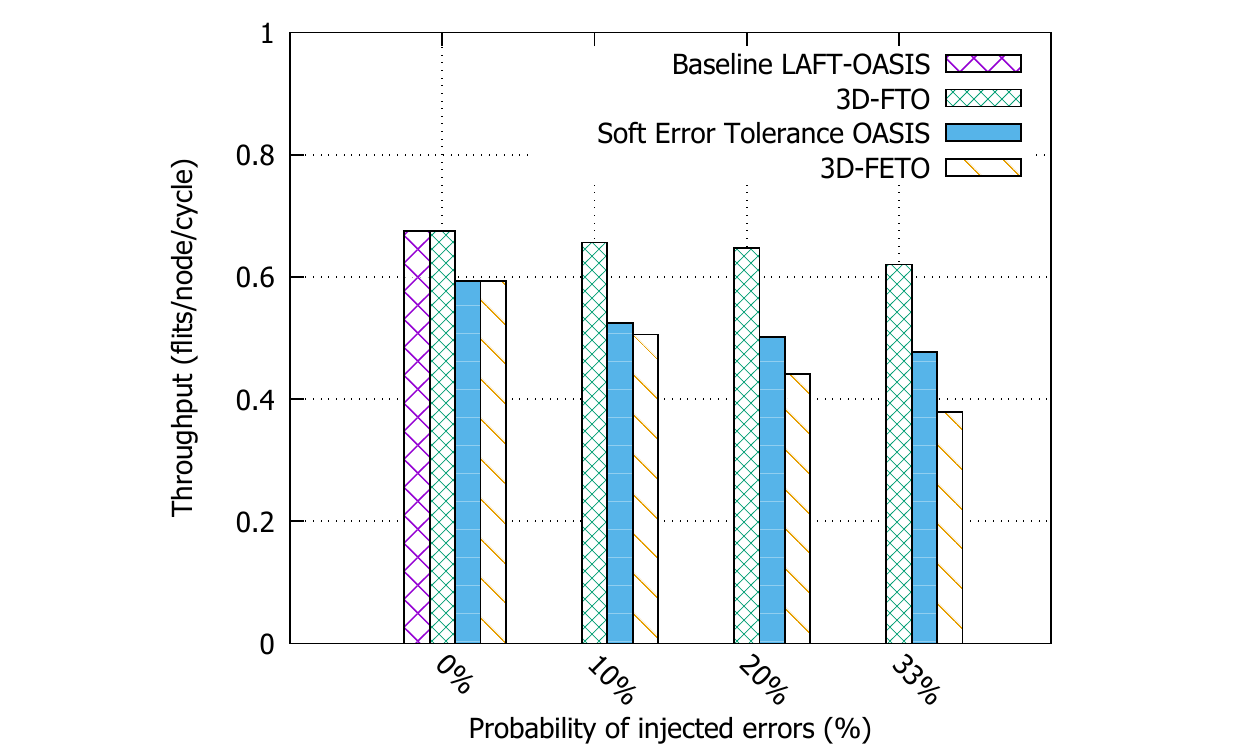}}
    \subfigure[Hotspot 10\%'s Throughput]{\includegraphics[width=0.3\linewidth]{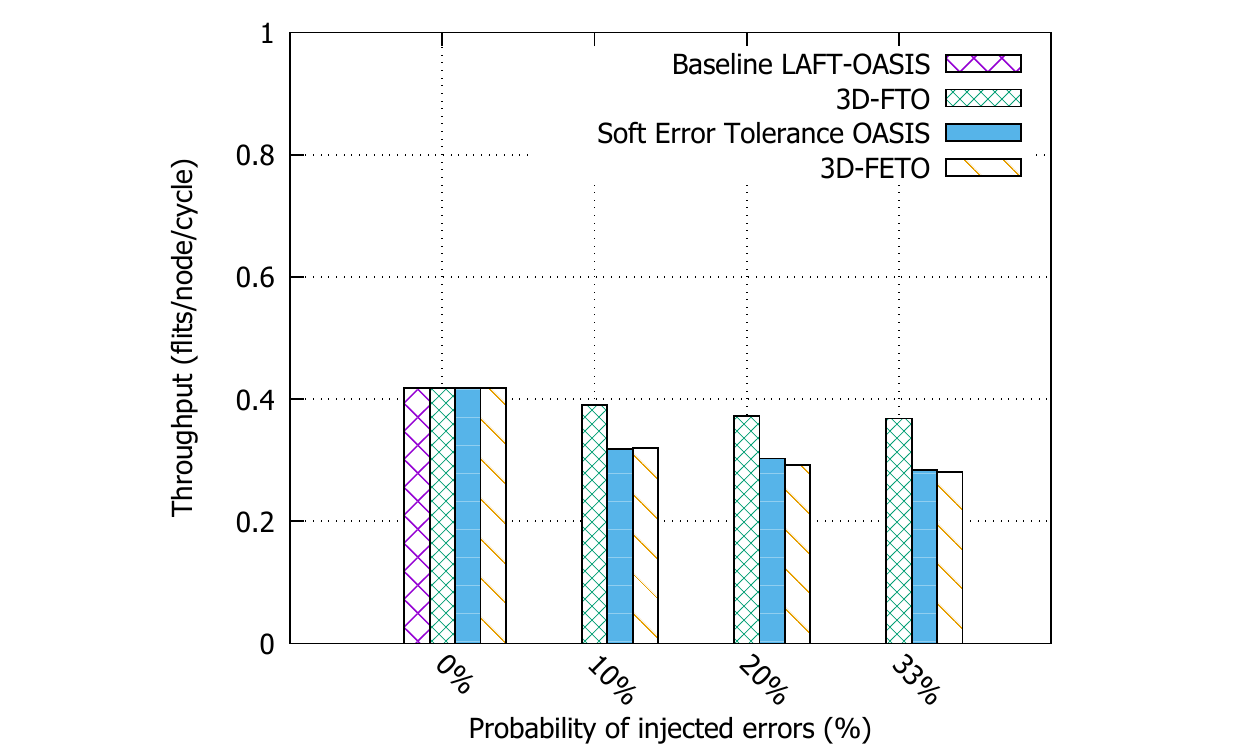}} 
    \caption{Average Packet Latency and Throughput  evaluation.}
    \label{fig:all-results}
    
\end{figure*}
Figure~\ref{fig:all-results} depicts the throughput evaluation with the adopted synthetic benchmarks. At 0\% error rate, 3D-FTO presents the best throughput which is matched to the capacity of the baseline LAFT-OASIS. The Soft Error Resilience OASIS and the proposed 3D-FETO have smaller throughput due to their soft error resilient mechanism. When the errors are injected into the system, we can observe a degradation in throughput. Thanks to the efficient hard fault tolerance scheme and the fault-tolerant routing algorithm, 3D-FTO at 33\% error-rate provides a slight decreased throughput: 40.18\%, 43.96\%, 43.55\% and 32.59\% for Transpose, Matrix, Uniform and Hotspot 10\% respectively. 
For the Soft Error Resilience OASIS, the system requires re-transmission in ARQ mechanism and re-execution in soft error mechanism. Therefore, the throughput is degraded due to extra clock cycles. The proposed 3D-FETO, which is a fusion of both hard fault tolerance and soft error resilience mechanisms, inherits both degradation. However, these systems provides the ability of error handling up to 33\% (the limitation of soft error mechanism).
\subsection{Complexity Evaluation}

Table~\ref{tb:fto-eval} illustrates the hardware complexity results of 3D-FETO in terms of area, power (static, dynamic, and total), and speed. In the hard fault tolerance router (3D-FTO), the area and power consumption overheads have increased by 1.01\% and 25.65\%, respectively. The maximum speed has also slightly decreased. On the other hand, our soft error handling mechanism adds seven ARQ buffers and some combinational logics which increase the area and power consumption significantly. However, the proposed 3D-FETO model introduces 7.50\% and 3.73\% extra area and power consumption, respectively, when compared to soft error resilience model. In comparison to the baseline model, 3D-FETO increases the area and power consumption by 56.39\% and 112.10\%, respectively, while the maximum speed decreases by 33.70\%.
\begin{table}[htbp]
    \scriptsize
    \begin{center}
        \caption{Simulation Configuration.}
        \label{tab:sim-conf}
        \begin{tabular}{|c|c||c|}
            \hline
            \multicolumn{2}{|c||}{\textbf{Parameter/System}} & \textbf{Value}  \\ \hline \hline
            \multirow{8}{*}{Network Size ($z\times y\times x$)} & Matrix & $3\times6\times6$  \\
            & Transpose & $4\times4\times4$ \\
            & Uniform                 & $4\times4\times4$ \\
            & Hotspot 10\%            & $4\times4\times4$ \\
            & H.264                   & $3\times3\times3$ \\   
            & VPOD                    & $2\times2\times3$ \\
            & MWD                     & $3\times2\times2$ \\
            & PIP                     & $2\times2\times2$ \\     \hline
            
            & Matrix       & 10             \\
            Node's Delivered Packets   & Transpose     & 10             \\
            per transmission session & Uniform      & 128  \\
            & Hotspot 10\% & 128  \\  \hline
            
            & H264     & 8,400     \\
            Network's Delivered Packets & VPOD     & 3,494     \\
            per transmission session  & MWD      &  1,120    \\
            & PIP      &  512      \\    \hline
            
            \multirow{2}{*}{Packet's Size}  & Hotspot 10\%  &  10 flits + 10\% for hotspot  \\ 
            & Others   &  10 flits   \\ \hline
            \multicolumn{2}{|c||}{Flits Size}    & 44 bits     \\ \hline
            \multicolumn{2}{|c||}{Header Size}   & 14 bits     \\ \hline
            \multirow{2}{*}{Payload Bit}  & Baseline, 3D-FTO  & 30 bits   \\ 
            & SER, 3D-FETO      &  18 bits  \\ \hline
            \multirow{2}{*}{Parity Bit}   & Baseline, 3D-FTO  & 0 bits    \\ 
            & SER, 3D-FETO      &  12 bits  \\ \hline
            
            \multicolumn{2}{|c||}{Buffer Depth}       & 4               \\ \hline

        \end{tabular}
    \end{center}
\end{table}        

Although our proposed models are penalized in terms of area, power consumption, and maximum frequency due to additional logics and registers that are necessary for fault handling mechanisms, they provide a full resiliency against a significant amount of soft errors and hard faults.    
\begin{table}[htbp]
    \begin{center}
        \scriptsize
        \caption{Hardware Complexity Evaluation.}
        \label{tb:fto-eval}
        \begin{tabular}{|l||c|c|c|c|c|}
            \hline
            & \textbf{Area} & \multicolumn{3}{c|}{\textbf{Power}} & \textbf{Speed}  \\
            \textbf{Model}             & ($\mu m^2$)   & \multicolumn{3}{c|}{(mW)}           & (Mhz)           \\ \cline{3-5}
            &               & Static   & Dynamic & Total          &                 \\ \hline \hline
            Baseline LAFT          & 18,873        &  5.1229  & 0.9429  & 6.0658         & 925.28          \\
            3D-FTO                 & 19,143        &  6.4280  & 1.1939  & 7.6219         & 909.09          \\
            Soft Error Resilience  & 27,457        &  9.7314  & 2.6710  & 12.4024        & 625.00          \\
            3D-FETO               & 29,516        &  10.0819 & 2.7839  & 12.8658        & 613.50          \\ \hline
        \end{tabular}
    \end{center}
\end{table} 

\section{Conclusion and Future Work}  \label{sec:concl}
In this paper, we proposed a comprehensive fault tolerant 3D-Network-on-Chip (3D-NoC) system architecture for highly-reliable many-core Systems-on-Chips (SoCs), named 3D-FETO. The proposed system is based on two approaches. First, a comprehensive mechanism to handle both soft error and hard faults in a 3D-NoC router is proposed.  
In the second approach, the system can support detection, diagnosis and recovery technique which makes it independent of any complex and costly testing mechanisms commonly found in conventional systems. 
Through extensive evaluation, we showed that the proposed 3D-FETO was able to recover efficiently from a significant number of soft and hard errors at different fault-rates reaching up to 33\%. 
Despite the performance degradation and hardware complexity penalty, we still consider that this overhead is acceptable. 

\section*{Acknowledgment}
This project is partially supported by Competitive Research Funding (CRF), University of Aizu, Japan, Ref.P-5-12, and JSPS Kakenhi Research Grant, Japan, Ref.30453020.


\end{document}